\documentclass[aps,prl,twocolumn,showpacs,amsmath,amssymb,superscriptaddress]{revtex4}

\usepackage{graphicx}
\usepackage{dcolumn}
\usepackage{bm}
\usepackage{xspace}

\graphicspath{{figures/}}

\newcommand{\nue}       {$\nu_{e}$\xspace}
\newcommand{\nuebar}    {$\bar{\nu_{e}}$\xspace}
\newcommand{\numu}      {$\nu_{\mu}$\xspace}
\newcommand{\nutau}     {$\nu_{\tau}$\xspace}

\newcommand{\mue}       {$\nu_{\mu} \rightarrow \nu_{e}$\xspace}
\newcommand{\mux}       {$\nu_{\mu} \rightarrow \nu_{x}$\xspace}
\newcommand{\mutau}     {$\nu_{\mu} \rightarrow \nu_{\tau}$\xspace}
\newcommand{\pizero}    {$\pi^{0}$\xspace}
\newcommand{\pizerogg}  {$\pi^{0}\to\gamma\gamma$\xspace}

\newcommand{\ssttmue}   {$\sin^2 2 \theta_{{\mu}e}$\xspace}
\newcommand{\ssttot}    {$\sin^2 2 \theta_{13}$\xspace}
\newcommand{\thetaot}   {$\theta_{13}$\xspace}
\newcommand{\dms}       {$\Delta m^2$\xspace}
\newcommand{\dmsmue}    {$\Delta m^2_{{\mu}e}$\xspace}

\newcommand{\dmsot}     {$\Delta m^2_{13}$\xspace}
\newcommand{\minv}      {$M_{inv}$\xspace}
\newcommand{\amome}     {$E_{e}$\xspace}

\begin{document}

\preprint{K2K/NUE05}

\title{Improved search for \mue oscillation in a long-baseline accelerator experiment}

\newcommand{\BCN}{\affiliation{Institut de Fisica d'Altes Energies, Universitat Autonoma de Barcelona, E-08193 Bellaterra (Barcelona), Spain}}
\newcommand{\BU}{\affiliation{Department of Physics, Boston University, Boston, Massachusetts 02215, USA}}
\newcommand{\UBC}{\affiliation{Department of Physics \& Astronomy, University of British Columbia, Vancouver, British Columbia V6T 1Z1, Canada}}
\newcommand{\UCI}{\affiliation{Department of Physics and Astronomy, University of California, Irvine, Irvine, California 92697-4575, USA}}
\newcommand{\SACLAY}{\affiliation{DAPNIA, CEA Saclay, 91191 Gif-sur-Yvette Cedex, France}}
\newcommand{\CNU}{\affiliation{Department of Physics, Chonnam National University, Kwangju 500-757, Korea}}
\newcommand{\DU}{\affiliation{Department of Physics, Dongshin University, Naju 520-714, Korea}}
\newcommand{\DUKE}{\affiliation{Department of Physics, Duke University, Durham, North Carolina 27708, USA}}
\newcommand{\GENEVA}{\affiliation{DPNC, Section de Physique, University of Geneva, CH1211, Geneva 4, Switzerland}}
\newcommand{\UH}{\affiliation{Department of Physics and Astronomy, University of Hawaii, Honolulu, Hawaii 96822, USA}}
\newcommand{\KEK}{\affiliation{High Energy Accelerator Research Organization(KEK), Tsukuba, Ibaraki 305-0801, Japan}}
\newcommand{\HIR}{\affiliation{Graduate School of Advanced Sciences of Matter, Hiroshima University, Higashi-Hiroshima, Hiroshima 739-8530, Japan}}
\newcommand{\INR}{\affiliation{Institute for Nuclear Research, Moscow 117312, Russia}}
\newcommand{\KOBE}{\affiliation{Kobe University, Kobe, Hyogo 657-8501, Japan}}
\newcommand{\KOR}{\affiliation{Department of Physics, Korea University, Seoul 136-701, Korea}}
\newcommand{\KYO}{\affiliation{Department of Physics, Kyoto University, Kyoto 606-8502, Japan}}
\newcommand{\LSU}{\affiliation{Department of Physics and Astronomy, Louisiana State University, Baton Rouge, Louisiana 70803-4001, USA}}
\newcommand{\MIT}{\affiliation{Department of Physics, Massachusetts Institute of Technology, Cambridge, Massachusetts 02139, USA}}
\newcommand{\MIYAGI}{\affiliation{Department of Physics, Miyagi University of Education, Sendai 980-0845, Japan}}
\newcommand{\NIIGATA}{\affiliation{Department of Physics, Niigata University, Niigata, Niigata 950-2181, Japan}}
\newcommand{\OKAYAMA}{\affiliation{Department of Physics, Okayama University, Okayama, Okayama 700-8530, Japan}}
\newcommand{\OSAKA}{\affiliation{Department of Physics, Osaka University, Toyonaka, Osaka 560-0043, Japan}}
\newcommand{\ROME}{\affiliation{University of Rome La Sapienza and INFN, I-000185 Rome, Italy}}
\newcommand{\SNU}{\affiliation{Department of Physics, Seoul National University, Seoul 151-747, Korea}}
\newcommand{\SOLTAN}{\affiliation{A.~Soltan Institute for Nuclear Studies, 00-681 Warsaw, Poland}}
\newcommand{\TOHOKU}{\affiliation{Research Center for Neutrino Science, Tohoku University, Sendai, Miyagi 980-8578, Japan}}
\newcommand{\SB}{\affiliation{Department of Physics and Astronomy, State University of New York, Stony Brook, New York 11794-3800, USA}}
\newcommand{\TUS}{\affiliation{Department of Physics, Tokyo University of Science, Noda, Chiba 278-0022, Japan}}
\newcommand{\KAM}{\affiliation{Kamioka Observatory, Institute for Cosmic Ray Research, University of Tokyo, Kamioka, Gifu 506-1205, Japan}}
\newcommand{\RCCN}{\affiliation{Research Center for Cosmic Neutrinos, Institute for Cosmic Ray Research, University of Tokyo, Kashiwa, Chiba 277-8582, Japan}}
\newcommand{\TRIUMF}{\affiliation{TRIUMF, Vancouver, British Columbia V6T 2A3, Canada}}
\newcommand{\VAL}{\affiliation{Instituto de F\'{i}sica Corpuscular, E-46071 Valencia, Spain}}
\newcommand{\UW}{\affiliation{Department of Physics, University of Washington, Seattle, Washington 98195-1560, USA}}
\newcommand{\WARSAW}{\affiliation{Institute of Experimental Physics, Warsaw University, 00-681 Warsaw, Poland}}

\BCN
\BU
\UBC
\UCI
\SACLAY
\CNU
\DU
\DUKE
\GENEVA
\UH
\KEK
\HIR
\INR
\KOBE
\KOR
\KYO
\LSU
\MIT
\MIYAGI
\NIIGATA
\OKAYAMA
\OSAKA
\ROME
\SNU
\SOLTAN
\TOHOKU
\SB
\TUS
\KAM
\RCCN
\TRIUMF
\VAL
\UW
\WARSAW

\author{S.~Yamamoto}\KYO 
\author{J.~Zalipska}\SOLTAN
\author{E.~Aliu}\BCN                
\author{S.~Andringa}\BCN 
\author{S.~Aoki}\KOBE 
\author{J.~Argyriades}\SACLAY 
\author{K.~Asakura}\KOBE 
\author{R.~Ashie}\KAM 
\author{F.~Berghaus}\UBC
\author{H.~Berns}\UW 
% ...
\collaboration{The K2K Collaboration}\noaffiliation

\date{\today}

\begin{abstract}
We performed an improved search for \mue oscillation with the KEK to Kamioka (K2K) long-baseline neutrino oscillation experiment, 
using the full data sample of $9.2 \times 10^{19}$\xspace protons on target.
No evidence for a \nue appearance signal was found, and
we set bounds on the \mue oscillation parameters. 
At \dms = $2.8 \times 10^{-3}~\mathrm{eV}^2$, 
the best fit value of the K2K \numu disappearance analysis,
we set an upper limit of \ssttmue $<$ 0.13 at 90\% confidence level.
\end{abstract}

\maketitle

{\itshape Introduction.}---
We describe a search for $\nu_e$ appearance in a beam of $\nu_\mu$, which is the signature of a non-zero value of the unknown neutrino mixing parameter $\theta_{13}$.
In the current picture of neutrino oscillations, three flavors of neutrinos are related to three mass states by the Maki-Nakagawa-Sakata matrix\cite{Maki:1962mu}. 
The mixing can be described by two $\Delta m^2$ parameters ($\Delta m_{\rm atm}^2, \Delta m_{\odot}^2$), three mixing angles ($\theta_{23}\sim \theta_{{\rm atm}}$, $\theta_{12}\sim \theta_{\odot}$, and $\theta_{13}$) and a CP violating phase ($\delta$). 
Atmospheric $\nu$ oscillations measured by several experiments~\cite{Ashie:2004mr,Ashie:2005ik,Sanchez:2003rb,Ambrosio:2003yz} and confirmed by the beam experiment~\cite{Aliu:2004sq} are well-described by $\nu_\mu \rightarrow \nu_\tau$ oscillations with parameters $\sin^2 2\theta_{\rm{atm}} > 0.92$ and $1.5\times 10^{-3} < \Delta m_{\rm{atm}}^2 < 3.4\times 10^{-3}~{\rm eV}^2$.
Solar $\nu_e \rightarrow \nu_{\mu,\tau}$ oscillations with parameters in the range $0.2 < \sin^2 \theta_{\odot} < 0.4$ and $7\times 10^{-5} < \Delta m_{\odot}^2 < 9\times 10^{-5}~{\rm eV}^2$ are also consistent with multiple observations \cite{Hosaka:2005um,Ahmed:2003kj}, and are confirmed by disappearance of reactor $\bar{\nu}_e$\cite{Araki:2004mb}.
As yet, very little is known about either $\theta_{13}$ or $\delta$, although lack of observed disappearance of reactor $\bar{\nu}_e$ over a few km baseline\cite{Apollonio:2002gd} has shown that $\theta_{13}$ must be smaller than 12$^\circ$ at the $\Delta m^2_{\rm atm}$ region reported by K2K\cite{Aliu:2004sq}.  
In a 2-flavor approximation, the probability of appearance of $\nu_e$ in a beam of $\nu_\mu$ is given by
\begin{equation*}\label{eqn1}
P(\nu_\mu \rightarrow \nu_e) = \sin^2 2\theta_{\mu e} \sin^2 (1.27 \Delta m^2_{\mu e} L/E), 
\end{equation*}
where $L$ is the baseline in km, $E$ is the neutrino energy in GeV and \dmsmue is in $\mathrm{eV^2}$.
For the case that $\Delta m^2_{23}\sim \Delta m^2_{\rm{atm}} \gg \Delta m^2_{12} \sim \Delta m^2_{\odot}$, and $\sin^2 2 \theta_{\rm{atm}} \sim 1$, we can take $\Delta m^2_{\mu e} \sim \Delta m^2_{\rm{atm}}$ and $\sin^2 2\theta_{\mu e} \sim \sin^2 \theta_{23} \sin^2 2\theta_{13} \sim \frac{1}{2} \sin^2 2 \theta_{13}.$

{\itshape Experimental apparatus and data sample.}---
The KEK to Kamioka (K2K) long-baseline neutrino experiment comprises a 98\% pure $\nu_\mu$ beam with mean energy of 1.3 GeV, created at KEK's proton synchrotron and sent 250~km to the Super-Kamiokande (SK) detector\cite{Fukuda:2002uc}. 
The beam is created by colliding primary protons of 12.9~GeV/c on an aluminum target, focusing the resulting secondary pions by two electromagnetic horns, and letting pions decay in a decay pipe.
The contamination of $\nu_e$ in the beam (beam \nue) is $\sim 1\%$ at KEK site.
Near neutrino detectors are employed at a 300~m baseline to measure the beam.
The near detector complex includes fine-grained detectors (FGD) and a 1 kiloton water Cherenkov detector (1KT). 
More details of the subdetectors are found in references~\cite{Suzuki:2000nj,Nitta:2004nt,Ishii:2001sj}. 

There were two running periods in K2K: K2K-I [$4.8\times 10^{19}$ protons on target (POT)] corresponds to dates for which SK was instrumented with 11,146 inner detector (ID) photomultiplier tubes (PMTs).
K2K-II ($4.4 \times 10^{19}$ POT) corresponds to dates after SK was rebuilt with 47\% ID PMT density following an accident. 
The K2K-I period is divided into K2K-Ia (June~1999) and K2K-Ib (November~1999 to July~2001).
For K2K-Ia, the horn current and target diameter were different from the subsequent phases of the experiment, and the differences of neutrino fluxes and systematic uncertainties are taken into account. 
For K2K-II (January~2003 to November~2004), the lead-glass calorimeter (LG) in FGD was replaced by a fully-active scintillator detector (SciBar).
We have analyzed the entire K2K data sample which corresponds to $9.2 \times 10^{19}$ POT in total.
The statistics are almost the double of our previous published results of K2K-I \cite{Ahn:2004te}.
The results in this paper are obtained with a revised signal selection and an improved sensitivity.

{\itshape Event selection.}---
Our search for \mue signal is based on detection of charged current quasi elastic (CC-QE) interaction of \nue in an oxygen nucleus: $\nu_e + n \to e + p$.
Typically the momentum of a recoil proton from CC-QE interaction is below Cherenkov threshold in the water. 
Therefore only the electron, visible as a single-ring, shower-type event, is the signature of $\nu_e$ appearance.

Beam neutrino events in SK are selected based on timing information from the global positioning system (GPS) to reject cosmic-ray and atmospheric neutrino background.
We start with the 112 fully-contained events in the 22.5 kiloton fiducial volume (FCFV) for K2K-I+II.
The definition of the electron signal candidates is given in Ref.~\cite{Ahn:2004te}.
The signal selection begins by identifying an electron ring candidate based on the shape of Cherenkov ring pattern and opening angle.
Next, signal candidates are required to have electron-equivalent energy (visible energy) above $100~\mathrm{MeV}$ to remove misidentified charged pions via \numu-CC interactions and electrons from muon decays in which muons are below Cherenkov threshold.
Finally, we require that no electron is followed by a muon decay within a 30 $\mu$sec time window. 

Near the limit of \thetaot\cite{Apollonio:2002gd} at the $\Delta m^2_{\rm atm}$ region (\ssttmue $\sim 0.05$, $\Delta m^2 \sim 2.8 \times 10^{-3}~\mathrm{eV^2}$), we obtain an oscillation probability for \nue appearance in K2K of $\sim 10^{-2}$ , which gives an expectation of a few \nue signal events in our entire sample.
Because this expected appearance signal is so small, precise understanding of background and its reduction are crucial.
One of the background sources is beam \nue.
This is responsible for $13\%$ of background events after the selection criteria described above. 
The remaining $87\%$ originate from beam \numu; this includes \numu-CC and neutral current (NC) interactions of \numu and \nutau, where \nutau appears as a result of \numu oscillation. 
This \numu-originated background is dominated by \pizero produced via NC interaction. 
A single \pizero decaying into two gamma rays can be classified by the standard atmospheric neutrino reconstruction procedure\cite{Ashie:2005ik} as a single-ring showering event, when one gamma ray is not reconstructed, due to highly asymmetric energies or a small opening angle between the two gamma rays.     

For the current analysis, we enhance the \pizero reconstruction capability with a newly introduced  algorithm for electron candidate events.
A second gamma-ray ring candidate is reconstructed by comparison of the observed charge and expected light patterns calculated under the assumption of two showering rings.
The direction of the ring reconstructed by the standard procedure, visible energy and vertex information are used as input.
We use the invariant mass of \pizerogg (\minv) from the momenta of two potential gamma-ray rings. 
Among single electron candidates, the \pizero background events have a value of \minv close to the \pizero mass.
In contrast, for a \nue signal event, the reconstructed energy of the second fake ring tends to be small after sharing the visible energy with the two-ring configuration.
Therefore \minv has a dependence on the energy of the electron candidate ring(\amome).
 
\begin{figure}[bp!]
  \begin{center}
    \includegraphics[width=\linewidth]{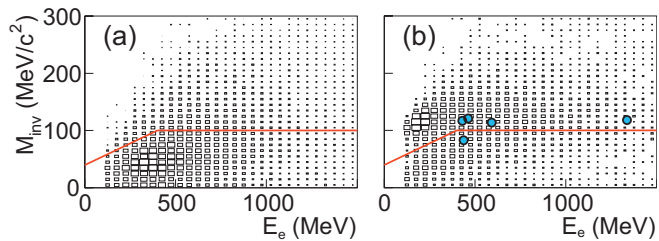}
    \caption{Distribution of candidate events in the \minv-\amome space. 
             The selection boundary is indicated by a line with a bend for (a) signal MC events and (b) \numu-originated background MC events. 
             Remaining events before the \pizero rejection in data (circles) are also shown for K2K-I and K2K-II combined.
             Signal MC distributions are obtained under the assumption of \ssttmue = 0.05 and \dms = $2.8 \times 10^{-3}~\mathrm{eV^2}$.}
    \label{figure:pi0cut}
  \end{center}
\end{figure}

For the final selection of candidate events, we exclude the region \minv $> 100~\mathrm{MeV/c^2}$.
Then a tighter cut is applied to the events with \amome below 400 MeV, where the resolution of \pizero mass is poor and the \nue signal shows small \minv.
The resultant \minv vs. \amome plot is shown in Fig. \ref{figure:pi0cut}.
The signal region is optimized to maximize statistical significance assuming (\ssttmue, \dms) = ($0.05, 2.8 \times 10^{-3} \mathrm{eV}^2$)\xspace.
With this requirement, 70\% of \numu-originated background in electron candidate events is suppressed, and the overall efficiency for selection of \nue via CC interactions for K2K-I and K2K-II is 47\% and 51\%, respectively. 
This selection efficiency of K2K-II is a little larger than K2K-I, however, the \numu background expectation is also larger for K2K-II.
The reduction of the data and background expectations according to the Monte Carlo (MC) simulation is summarized in Table~\ref{table:reduction}. 
After all selection cuts, we obtain one signal candidate in the data; four of five remaining before the final selection are rejected as \pizero-like.
(The single signal candidate of K2K-I reported in Ref.~\cite{Ahn:2004te} is rejected by the new cuts.)
This rejection capability is as expected, and a signal candidate is consistent with background expectation.
A visual examination reveals the surviving candidate to be an event with more than two rings.
It is consistent with a multihadron production event; these make up 14\% of the background according to the MC simulation.

\begin{table}[bp!]
  \caption{Reduction of events for \mue search at SK.
           The first column lists each selection requirement. 
           The others give, for each selection, the number of observed events, \numu background in the case of no oscillation, and beam \nue, for K2K-I and K2K-II, respectively.}
           
  \label{table:reduction}
  \begin{center}
    \begin{tabular}{lcccccc}
      \hline
       & \multicolumn{3}{c}{K2K-I} & \multicolumn{3}{c}{K2K-II}\\
       & Data & \numu & Beam-\nue & Data & \numu & Beam-\nue \\
      \hline
      FCFV               & 55 & 81.1 & 0.8 & 57 & 77.4 & 0.9 \\
      single ring        & 33 & 50.1 & 0.5 & 34 & 49.4 & 0.5 \\
      shower like        &  3 &  2.7 & 0.4 &  5 &  3.2 & 0.4 \\
      visible energy cut &  2 &  2.5 & 0.4 &  5 &  2.9 & 0.4 \\
      no decay electron  &  1 &  1.9 & 0.3 &  4 &  2.2 & 0.4 \\
      non-$\pi^0$ like   &  0 &  0.6 & 0.2 &  1 &  0.7 & 0.2 \\
      \hline
    \end{tabular}
  \end{center}
\end{table}

{\itshape Background expectation.}---
In our MC simulation, neutrino interactions with oxygen nuclei are simulated as in Ref.\cite{Hayato:2002sd}.
The interaction models used in this analysis are the same as the \mux oscillation analysis \cite{Aliu:2004sq} except that we set the cross section for CC coherent pion production to zero, based on Ref. \cite{Hasegawa:2005td}.

The \numu energy spectrum and normalization are derived from measurements at the near detectors.
The far/near $\nu$ spectrum ratio used to extrapolate near detector measurements to SK is calculated using a beam MC simulation.
In the simulation, we employ the $\pi^+$ production cross-section measured by the HARP experiment\cite{Catanesi:2005rc}.
This spectrum ratio is validated by in situ measurements of pions from the aluminum target\cite{Ahn:2001cq}.
With the full data set the resulting flux is consistent with the previous result\cite{Aliu:2004sq}.
The normalization of the \numu events is determined by the 1KT data.  
The expected total number of FCFV events at SK without oscillation for K2K-I and K2K-II is 81.1 and 77.4, respectively.

The number of background events is estimated to be $0.8~(0.9)$ for K2K-I (K2K-II); 0.6 (0.7) event is originated from \numu and the remaining 0.2~(0.2) is beam \nue.
The background events are dominated by NC interactions and are consequently nearly insensitive to \numu oscillated to \nutau;
\nutau oscillation affects the background estimation by $\sim5\%$
at the oscillation parameters of the K2K result \cite{Aliu:2004sq}.

{\itshape Systematic uncertainties.}---
The various contributions to the systematic uncertainty in \numu-originated background are summarized in Table~\ref{table:numu_uncertainties}. 
Most of the error sources are related to \pizero background in NC interactions.
The fraction of NC single \pizero (NC1\pizero) production in 
the total of background events in SK is about 70\%;
the 1KT detector data were used for 
understanding its cross section in water.
The 1KT measurement of the cross section ratio of NC1\pizero production to 
CC interaction (NC1\pizero/CC) gives 
$0.064\pm 0.001(stat.) \pm 0.007(sys.)$ \cite{Nakayama:2004dp},
and our MC simulation predicts the ratio to be 0.065.
Taking into account this measurement and the difference of the central values, 
we apply an uncertainty of 12\% to the NC1\pizero/CC ratio.
For the composition of NC background other than NC1\pizero, 
an uncertainty of 20\% in the ratio of NC cross section to CC is quoted 
as in Ref.\cite{Monsay:1978}.
For coherent NC \pizero production, we account for the difference between the zero coherent pion production case and the model of Rein and Sehgal \cite{Rein:1982pf}. 
Varying these cross sections by their errors, the contributions to the expected number of events are estimated.
In addition, final state interactions of nucleons and mesons inside of nuclear matter
could substantially alter \pizero momenta.
This effect is estimated to be 8\% on the \numu-originated background, considering the differences in the \pizero spectrum between 1KT data and MC events.

Systematic uncertainty related to the \pizero rejection has been evaluated based on the measurement of atmospheric neutrinos in SK. 
For this purpose we select two data samples which include an unreconstructed gamma ray from \pizero decays in the final state. One of these is a sample of single electron events; the other is a sample of events with one muon-like and one electron-like ring reconstructed. 
Using both samples, an uncertainty of $19\%$ is estimated in the \minv cut in order to account for the effect of possible reconstruction biases which affect the \minv distribution.
Water properties also affect the \pizero reconstruction and associated errors are estimated using cosmic ray muons.
The light attenuation has been measured during the entire period of operation of the SK detector.
The absorption and scattering lengths were varied within their ranges accordingly, $\pm 20\%(\pm 15\%)$ for SK-I (SK-II), which results in systematic effects of $\pm11\%$ ($\pm6\%$) for K2K-I (K2K-II).

\begin{table}[tbp!]
  \caption{Systematic uncertainties [\%] in the expectation of \numu-originated  background.
  When estimating the total uncertainty, the correlations between the neutrino fluxes and the cross sections are taken into account.}
  \label{table:numu_uncertainties}
  \begin{center}
    \begin{tabular}{lccc}
      \hline
       & \multicolumn{2}{c}{K2K-I} & K2K-II \\
       & (Ia) & (Ib) &  \\
      \hline
      NC1\pizero/CC ratio          & $\pm8$         & $^{+6}_{-7}$      & $^{+6}_{-7}$ \\
      NC/CC ratio (non-NC1\pizero) & $\pm4$         & $\pm3$            & $\pm3$ \\
      \pizero energy spectrum      & \multicolumn{2}{c}{$\pm8$}         & $\pm8$ \\
      coherent \pizero model       & \multicolumn{2}{c}{$^{+3}_{-10}$}  & $^{+3}_{-10}$ \\
      \pizero mass cut             & \multicolumn{2}{c}{$^{+19}_{-18}$} & $^{+19}_{-17}$ \\
      water properties             & \multicolumn{2}{c}{$\pm11$}        & $\pm6$ \\
      neutrino flux at SK          & $^{+20}_{-17}$ & $\pm6$            & $\pm6$\\
      non-QE/QE ratio              & $^{+2}_{-2}$   & $^{+1}_{-1}$      & $^{+1}_{-1}$ \\
      detector efficiency          & \multicolumn{2}{c}{$\pm6$}         & $\pm6$ \\ 
      single electron selection    & \multicolumn{2}{c}{$^{+5}_{-7}$}   & $^{+7}_{-8}$ \\
      \hline
      total & $^{+33}_{-32}$ & $^{+37}_{-26}$ & $^{+39}_{-24}$\\
      \hline
    \end{tabular}
  \end{center}
\end{table}

For the beam-\nue contamination, the expectation is derived from the \nue/\numu flux ratio at SK with the beam MC simulation and \numu flux extrapolation from the near detectors to SK. 
The beam MC expectation has been verified by measurements of \nue/\numu interaction ratio at the FGD complex.
The measurement by SciBar\cite{K2K:2006}, covering the electron energy above $0.5~\mathrm{GeV}$, gives this ratio as $1.6 \pm 0.3 (\mathrm{\it stat.}) \pm 0.2 (\mathrm{\it syst.})\%$.
The ratio measured by LG\cite{Ahn:2004te} for the electron energy above $1~\mathrm{GeV}$ is consistent with that of SciBar, and both show agreement with the beam MC prediction of $1.3\%$.
The systematic uncertainty in the number of beam-\nue background is dominated by our understanding of pion and kaon production in proton collisions on the aluminum target;
\nue contamination via muon and kaon decays is taken into account, where muons are emitted by positive pion decays.
Uncertainties in the beam-\nue background of 14\% and 16\% are estimated for pion and kaon production, respectively.
Incorporating other possible contributions such as the single electron selection, we quote an uncertainty of $^{+32}_{-21}\%$ in total.

For the number of appearance signal events, the total estimated uncertainty is 15\% with dominant sources coming from the \numu energy spectrum and the cross section ratio of CC interactions other than CC-QE (non-QE) to CC-QE measured by the near detectors.

{\itshape Limits on mixing parameters.}---
To set an excluded parameter region in a 2-flavor oscillation model, we adopt a confidence interval construction with the Poisson distribution using the expected number of events and the observed number; 
these values are $1.7^{+0.6}_{-0.4}$ (in the case of no oscillation) and 1, respectively.
The expected number of events $\mathcal{N}_{obs}$ can be represented by the sum of two background components and an appearance signal as
\begin{align*}
  \mathcal{N}_{obs} = &
  \mathcal{B}_{\nu_{\mu}}(\Delta m^2_{\mu e}, \sin^2 2\theta_{\mu e}) + 
  \mathcal{B}_{beam-\nu_{e}} \\
  & + \mathcal{S}_{\nu_{e}}(\Delta m^2_{\mu e}, \sin^2 2\theta_{\mu e}),
\end{align*}
where $\mathcal{B}_{\nu_{\mu}}$\xspace is the number of electron candidate events induced by \numu or oscillated to \nutau, and $\mathcal{B}_{beam-\nu_{e}}$\xspace is that of beam-\nue.
The number of \mue appearance signal, $\mathcal{S}_{\nu_{e}}$, and $\mathcal{B}_{\nu_\mu}$ depend on the probability of \mue oscillation.
For the contribution of \mutau oscillation to $\mathcal{B}_{\nu_\mu}$, we assume the best-fit parameters of the K2K \numu disappearance analysis\cite{Aliu:2004sq}.
In the calculation of the upper limit on \ssttmue, the effects of systematic uncertainties are incorporated into the probability densities and the unified ordering prescription of Feldman and Cousins \cite{Feldman:1997qc} is applied.

\begin{figure}[tp!]
  \begin{center}
    \includegraphics[height=50mm]{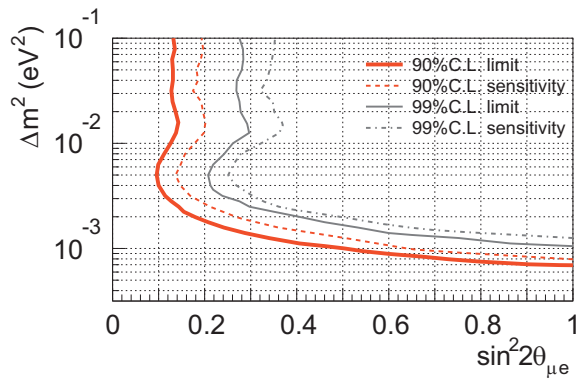}
    \caption{The upper bound on \mue oscillation parameters at 90\% and 99\% C.L.
    The sensitivities of the K2K experiment for each C.L. are also indicated with dashed lines.}
    \label{figure:contour_k2k1-2}
  \end{center}
\end{figure}

Figure \ref{figure:contour_k2k1-2} shows the upper bound on the oscillation parameters for two flavor mixing, at the  90\% and 99\% confidence level (C.L.).
Neutrino oscillation from \numu to \nue is excluded at 90\% C.L.  in \ssttmue $>$ 0.13 at \dmsmue = $2.8 \times 10^{-3} \mathrm{eV}^2$.

Reactor \nuebar experiments provide complementary results to a search for non-zero \thetaot with a \numu beam; 
currently the only result for \nue appearance mode is provided by K2K.
We note our resulting upper limit at 90\% C.L. is \ssttot $=0.26$ at \dmsot $=2.8\times10^{-3} \mathrm{eV}^2$ assuming \ssttmue $= \frac{1}{2}$ \ssttot and \dmsmue $\sim$ \dmsot.
In the same \dms region, the most stringent limit from reactor experiments is \ssttot $=0.1$ by the CHOOZ experiment \cite{Apollonio:2002gd} and a weaker limit of 0.16 is reported by the Palo-Verde experiment \cite{Piepke:2002ju}, showing agreement with our result.

In summary, the K2K experiment finished taking data in November 2004.
Starting from June 1999, we accumulated data which corresponds to $9.2 \times 10^{19}$ POT.
Compared to the previous search\cite{Ahn:2004te}, we improved both the statistics and the rejection of \pizero backgrounds.
As a result, we find no evidence for neutrino oscillations in the \nue appearance mode.
A single electron candidate is consistent with background expectation.
We set an upper limit of \ssttmue $<$ 0.13 at 90\% C.L.  at the best-fit parameters of the \numu disappearance analysis.

\begin{acknowledgments}
We wish to thank the KEK and ICRR directorates for their strong support and encouragement.  
K2K is made possible by the inventiveness and the diligent efforts of the KEK-PS machine group and beam channel group.
We gratefully acknowledge the cooperation of the Kamioka Mining and Smelting Company. 
This work has been supported by the Ministry of Education, Culture, Sports, Science and Technology of the Government of Japan, the Japan Society for Promotion of Science, the U.S. Department of Energy, the Korea Research Foundation, the Korea Science and Engineering Foundation, NSERC Canada and Canada Foundation for Innovation, the Istituto Nazionale di Fisica Nucleare (Italy), the Spanish Ministry of Science and Technology, and Polish KBN grants: 1P03B08227 and 1P03B03826.
\end{acknowledgments}

\end{document}